\documentstyle[prb,aps,multicol,epsfig]{revtex}

\begin{document}

\draft

\title{The spin-orbit interaction as a source of \\
new spectral and transport properties in quasi-one-dimensional systems}

\author{A.~V.~Moroz and C.~H.~W.~Barnes}

\address{Cavendish Laboratory,
     University of Cambridge, \\
     Madingley Road,
     Cambridge CB3 0HE,
     United Kingdom}

\date{\today}

\maketitle

\begin{abstract}

We present an exact theoretical study of the effect of the spin-orbit (SO)
interaction on the band structure and low temperature transport in long
quasi-one-dimensional electron systems patterned in two-dimensional
electron gases in zero and weak magnetic fields. We reveal the
manifestations of the SO interaction which cannot in principle be observed
in higher dimensional systems.

\end{abstract}


\begin{multicols}{2}
\narrowtext

It is known that an electron moving in an electric field experiences not
only an electrostatic force but also a relativistic influence that is
referred to as the spin-orbit (SO) interaction (or spin-orbit
coupling~\cite{SO}). It is caused by Pauli coupling between the spin
moment of an electron and a magnetic field which appears in the rest frame
of the electron due to its motion in the electric field. The Hamiltonian of
the SO interaction has the form~\cite{SO}:

\begin{equation}
\hat H_{SO} = -\frac{\hbar}{(2M_0c)^2}\;{\bf E}({\bf R})
\left[\hat{\bbox{\sigma}}\times
\left\{\hat{\bf p}+\frac{e}{c}{\bf A}({\bf R})\right\}\right].
\label{H_SO}
\end{equation}
Here $M_0$ is the free electron mass, $\hat{\bf p}$ is the canonical
momentum operator,
$\hat{\bbox{\sigma}}$ is the Pauli matrices, ${\bf E}({\bf R})$ is the
electric field, ${\bf A}({\bf R})$ is a vector potential, and ${\bf R}$ is
a 3D position vector. Usually the Hamiltonian (\ref{H_SO}) results in a
spin-orientation dependence of the electron energy and/or wave functions.
This dependence can become important if electric fields acting on a system
of moving electrons are sufficiently strong.

One of the most promising solid-state nanostructures for the observation
of SO-interaction effects is the quasi-one-dimensional electron
system~\cite{Thornton,Kelly} (Q1DES) patterned in two-dimensional electron
gases (2DEG). In contrast with higher dimensional structures,
Q1DES have {\it three} independent sources of strong electric
fields: (i) crystal-field potential that is present in all
dimensionalities owing to the intermolecular forces; (ii) a quantum-well
potential~\cite{Kelly} that confines electrons to a 2D layer at the surface
of the crystal; (iii) a transverse (in-plane) electric potential that is
applied to squeeze the 2DEG into a quasi-one-dimensional
channel~\cite{Thornton,Kelly}. The strength of the in-plane potential
determines an effective width of a Q1DES that can be controlled by
changing the transverse voltage. In sufficiently narrow channels the
transverse electric field can be made comparable with the other two
electrostatic contributions.

The study of the SO interaction in Q1DES is interesting from
the standpoint of remarkable transport phenomena which they exhibit:
ballistic qiantisation of conductance~\cite{WW}; the 0.7 conductance
structure~\cite{TKT}; magnetic depopulation~\cite{vanWees}; and negative
magnetoresistance~\cite{NMR}. Since these phenomena depend on the
peculiarities of the energy spectrum of electrons, any new mechanism
leading to non-trivial changes in the spectrum (especially to those
involving the spin) may affect transport properties and thereby help their
understanding.

Earlier theoretical~\cite{Dress,Rashba,Eppenga,Edelstein} and
experimental~\cite{Stormer,Dorozhkin,Das,Luo,Nitta,Hassenkam} works on the
SO-related effects dealt mainly with 3D and 2D systems and did not touch on
aspects of the SO coupling in Q1DES. In this paper we present the results
of a theoretical analysis of the effect of the SO interaction on the energy
spectrum and low temperature (ballistic) conductance of a long Q1DES. Since
the crystal-field contribution to the SO interaction energy can be made
negligible in comparison with the quantum-well effect in a variety of
systems~\cite{Dorozhkin,Das,Luo,Nitta,Hassenkam}, we take into account two
sources of the SO coupling: the quantum-well confinement in the direction
perpendicular to the plane of the 2DEG and the confining electric potential
transverse to the 2DEG. We show that even if the SO coupling due to the
transverse potential is left out, the very presence of this potential
changes drastically the SO-interaction effects caused by the quantum-well
field in comparison with a purely 2D situation. In addition to this, the
contribution of the transverse potential to the SO coupling adds
interesting features to the electron energy spectrum and the conductance
which cannot be accounted for by simply renormalising the quantum-well
field. Also we find that relatively weak magnetic fields emphasise the
effects of the SO interaction in Q1DES.

A unique feature of semiconductor Q1DES is that their properties
can be varied significantly at the stage of design (via chemical
composition, band engineering, external fields, etc.). In
particular, it is possible to create systems with wide range of
carrier concentrations including values where the strength of
electron-electron interactions is relatively weak. On the other
hand, the strength of the SO coupling can simultaneously be
enhanced by, for example, increasing confining electric fields and
using materials with larger SO constants (InAs, PbTe, etc.). As a
result, experimental situations can be achieved when the SO
coupling becomes dominant. In this case it is reasonable to assume
that the electron-electron interaction does not remove SO effects
in the band structure of Q1DES and they can be studied within the
single-particle approximation. As regards the ballistic
conductance of a quantum wire, it has been proved not to be
renormalised by electron-electron interactions~\cite{Maslov}.
Based on these arguments, we consider here a free 2DEG within
one-band effective mass approximation~\cite{Ridley}. The
corresponding Hamiltonian has the form:

\begin{equation}
\hat H = \frac{1}{2M}\left(\hat{\bf p}+\frac{e}{c}{\bf A}\right)^2
+V({\bf R})+\frac{g}{2}\mu_B\left(\hat{\bbox{\sigma}}{\bf B}\right)
+\hat H_{SO}.
\label{H}
\end{equation}
Here $M$ is the effective electron mass, $g$ is the Land\'{e} $g$-factor,
and $\mu_B$ is the Bohr magneton. The vector potential ${\bf A}$ is chosen
in the Landau gauge ${\bf A}({\bf R})=Bx\,{\bf y}$ with a magnetic field
${\bf B}={\rm curl}\; {\bf A}=B{\bf z}$ being perpendicular to the 2DEG. In
line with Refs.~\cite{Laux,Drexler,Kardynal}, the {\it transverse confining
potential} $V({\bf R})$ is approximated by a parabola

\begin{equation}
V({\bf R}) \equiv V(x) = (M\omega^2/2)x^2,
\label{V}
\end{equation}
where $\omega$ controls the strength of the confining potential.

We assume that the SO Hamiltonian $\hat H_{SO}$ (\ref{H_SO}) in
Eq.~(\ref{H}) is formed by two contributions: $\hat H_{SO}=\hat
H_{SO}^\alpha+\hat H_{SO}^\beta$. The first one, $\hat H_{SO}^\alpha$,
arises from the quantum-well electric field that can reasonably be assumed
uniform and directed along the $z$-axis, so that $\hat H_{SO}^\alpha$ is
given by

\begin{equation}
\hat H_{SO}^\alpha = \frac{\alpha}{\hbar}
\left[\hat{\bbox{\sigma}}\times\left(\hat{\bf p}+\frac{e}{c}{\bf A}
\right)\right]_z.
\label{Ha}
\end{equation}
The SO-coupling constant $\alpha$ takes values $10^{-10}$ -- $10^{-9}$
eV$\times$cm for different systems~\cite{Rashba,Das,Luo,Nitta,Hassenkam}.
We will refer to this mechanism of the SO coupling as $\alpha$-coupling.

The second contribution $\hat H_{SO}^\beta$ to $\hat H_{SO}$ comes from the
in-plane electric field ${\bf E}=-\nabla_{{\bf R}} V=-M\omega^2{\bf x}$
caused by the transverse confining potential (\ref{V}):

\begin{equation}
\hat H_{SO}^\beta = \frac{\beta}{\hbar}
\frac{x}{l_\omega}
\left[\hat{\bbox{\sigma}}\times\left(\hat{\bf p}+\frac{e}{c}{\bf A}
\right)\right]_x,
\quad l_\omega=\sqrt{\hbar/M\omega}.
\label{Hb}
\end{equation}
By comparison with typical quantum-well and transverse electric fields the
SO-coupling constant $\beta$ in Eq.~(\ref{Hb}) can be roughly estimated as
at least $\beta \sim 0.1\,\alpha$. Moreover, in square quantum wells where
the value of $\alpha$ is considerably diminished~\cite{Hassenkam} the
constant $\beta$ may well compete with $\alpha$. We will call the SO
interaction arising from the transverse confining potential (\ref{V})
$\beta$-coupling.

To calculate the energy spectrum of electrons we must find eigenvalues $E$
of the Schr\"{o}dinger equation $\hat H\Psi=E\Psi$ where the wave function
$\Psi=\Psi({\bf R})=\left\{\Psi({\bf R})_\uparrow \; \Psi({\bf
R})_\downarrow\right\}$ is a two-component spinor. Since the Hamiltonian
$\hat H$ (\ref{H}) -- (\ref{Hb}) is translationally invariant in the
$y$-direction, the wave functions $\Psi_{\uparrow\downarrow}({\bf R})$ are
plane waves propagating along the $y$-axis, i.e.
$\Psi_{\uparrow\downarrow}({\bf
R})=\exp(ik_yy)\Phi_{\uparrow\downarrow}(x)$, and the longitudinal energy
is given by $E_y=\hbar^2 k_y^2/2M$, where $k_y$ is the longitudinal wave
number. The equations for $\Phi_{\uparrow\downarrow}(x)$ stem from the
Schr\"{o}dinger equation:

\begin{eqnarray}
&\Phi_{\uparrow\downarrow}^{\prime\prime}+
\left[\varepsilon_x \mp \gamma(l_\omega/l_B)^2
-a_{\uparrow\downarrow} t^2-b_{\uparrow\downarrow} t\right]
\Phi_{\uparrow\downarrow}(t)= \nonumber \\
&(l_\omega/l_\alpha)\left\{\pm\Phi_{\downarrow\uparrow}^\prime+
\left[(l_\omega/l_B)^2t+k_yl_\omega\right]
\Phi_{\downarrow\uparrow}(t)\right\},
\label{Phi}
\end{eqnarray}

\begin{eqnarray}
a_{\uparrow\downarrow}&=&1+(l_\omega/l_B)^4 \pm
(l_\omega/l_B)^2(l_\omega/l_\beta),
\label{a} \\
b_{\uparrow\downarrow}&=&2(k_yl_\omega)\left[(l_\omega/l_B)^2 \pm
(1/2)(l_\omega/l_\beta)\right],
\label{b}
\end{eqnarray}
where $l_B=\sqrt{c\hbar/eB}$ is the magnetic length and
$l_{\alpha(\beta)}=\hbar^2/2M\alpha(\beta)$ are typical spatial scales
associated with the $\alpha$- ($\beta$-) coupling. The quantities
$\varepsilon_x \equiv (k_xl_\omega)^2$ and $t=x/l_\omega$ are the
dimensionless transverse energy and coordinates,
$k_x^2=(2M/\hbar^2)E-k_y^2$, and $\gamma=(M/M_0)g/2$.

As opposed to all the other terms in Eq.~(\ref{H}), the operator $\hat
H_{SO}^\alpha$ (\ref{Ha}) is {\it non-diagonal} in the spin space.
Therefore, as long as the $\alpha$-coupling is finite (i.e. if
$l_\omega/l_\alpha \neq 0$), the equations (\ref{Phi}) are coupled to each
other. It is therefore natural that the behaviour of the transverse energy
$\varepsilon_x$ which is determined by Eqs.~(\ref{Phi}) crucially depends
on whether or not the $\alpha$-coupling is present in the system.

For zero $\alpha$-coupling ($l_\omega/l_\alpha=0$) Eqs.~(\ref{Phi})
decouple and reduce to analytically solvable Hermite equations. The
transverse energy is then given by

\begin{eqnarray}
\varepsilon_x^{\uparrow\downarrow}&=&
(2n+1)a_{\uparrow\downarrow}^{1/2} \pm \gamma(l_\omega/l_B)^2-
\nonumber \\
&&a_{\uparrow\downarrow}^{-1}\left[(l_\omega/l_B)^2 \mp
(1/2)(l_\omega/l_\beta)\right]^2(k_yl_\omega)^2,
\label{E0}
\end{eqnarray}
$n=0,1,2,\ldots$ and the wave functions $\phi_{\uparrow\downarrow}^n(t)$
form complete sets. The functions $\varepsilon_x^{\uparrow\downarrow}=
\varepsilon_x^{\uparrow\downarrow}(k_y)$ resemble well-known
magneto-electric parabolic subbands~\cite{Datta} with the only exception
that finite $\beta$-coupling brings in a spin-orientation dependence of the
subband curvature.

We now consider the case of finite $\alpha$-coupling ($l_\omega/l_\alpha
\neq 0$). We do not find that the coupled Eqs.~(\ref{Phi}) can
be solved in an explicitly analytical form. However, a strongly
convergent matrix form exists.  This is found by expanding {\it
each} unknown wave function $\Phi_\uparrow(t)$ and $\Phi_\downarrow(t)$ in
terms of {\it both} $\phi_\uparrow^n(t)$ and $\phi_\downarrow^n(t)$ and then
combining all {\it four} expansions obtained into a {\it closed}
linear homogeneous system of algebraic equations with respect to the
coefficients of one of the expansions. The exact spectrum $\varepsilon_x$
has been found numerically as zeros of the corresponding determinant as a
function of $k_y$ (see Ref.~\cite{MB} for more details for the
zero-magnetic-field case). The exploitation of the {\it four} expansions
has allowed us to avoid inversions of infinite matrices, while the
conveniently chosen bases have made the roots of the determinant rapidly
convergent.

Solid lines in Fig.~\ref{fig1}(a) present graphs of
$\varepsilon_x=\varepsilon_x(k_yl_\omega)$ for zero $\beta$-coupling
($l_\omega/l_\beta=0$) and zero magnetic field. Here we see two-fold spin
degeneracy of all quantum levels at $k_y=0$. Once $k_y$ becomes finite, the
SO interaction lifts this degeneracy producing an energy splitting
$\Delta\varepsilon_x=\varepsilon_x^\uparrow-\varepsilon_x^\downarrow \neq
0$ between electron states with different spin orientations. For small
$k_yl_\omega \lesssim 2$ this splitting is linear in $k_y$ and agrees with
results of both theoretical~\cite{Rashba,Bychkov} and
experimental~\cite{Das,Luo,Nitta} research on the SO-interaction effects
caused by the quantum-well field in 2D systems. However, in a purely 2D
geometry the {\it linear} splitting $\Delta\varepsilon_x \propto k_y$ is
known~\cite{Rashba} to be exact for {\it all} values of $k_y$, however
large. In contrast to this, the Q1DES dispersion curves start to diverge
from the linear behaviour at $k_yl_\omega \approx 2.5$ and eventually {\it
anticross} with an energy branch corresponding to the next higher (lower)
quantum number $n$.  This is a direct consequence of the
presence of the transverse confining potential (\ref{V}). Even though this
potential does not contribute to the SO interaction directly (because
$l_\omega/l_\beta=0$), it nevertheless strongly affects the other
(quantum-well) mechanism of the SO coupling. More specifically, in the
presence of the potential (\ref{V}) the transverse wave functions
$\Phi_{\uparrow\downarrow}(x)$ of the unperturbed system (i.e. with
$l_\omega/l_\alpha=0$) are no longer simple plane waves $\exp(ik_xx)$ (as
it is in a stritcly 2D situation) but become parabolic cylinder
functions~\cite{SO}. When the SO perturbation operator [the rhs' of
Eqs.~(\ref{Phi})] acts on these functions, it projects the $n$-th state
onto the $(n \pm 1)$-st states producing an effective {\it hybridisation}
of ``neighbouring'' states and therefore the {\it anticrossing} of the
energy branches in Fig.~\ref{fig1}(a) and the non-monotonic dependence
$\Delta\varepsilon_x(k_y)$ (see Ref.~\cite{MB} for more details).

\begin{figure}
\begin{center}
\epsfig{file=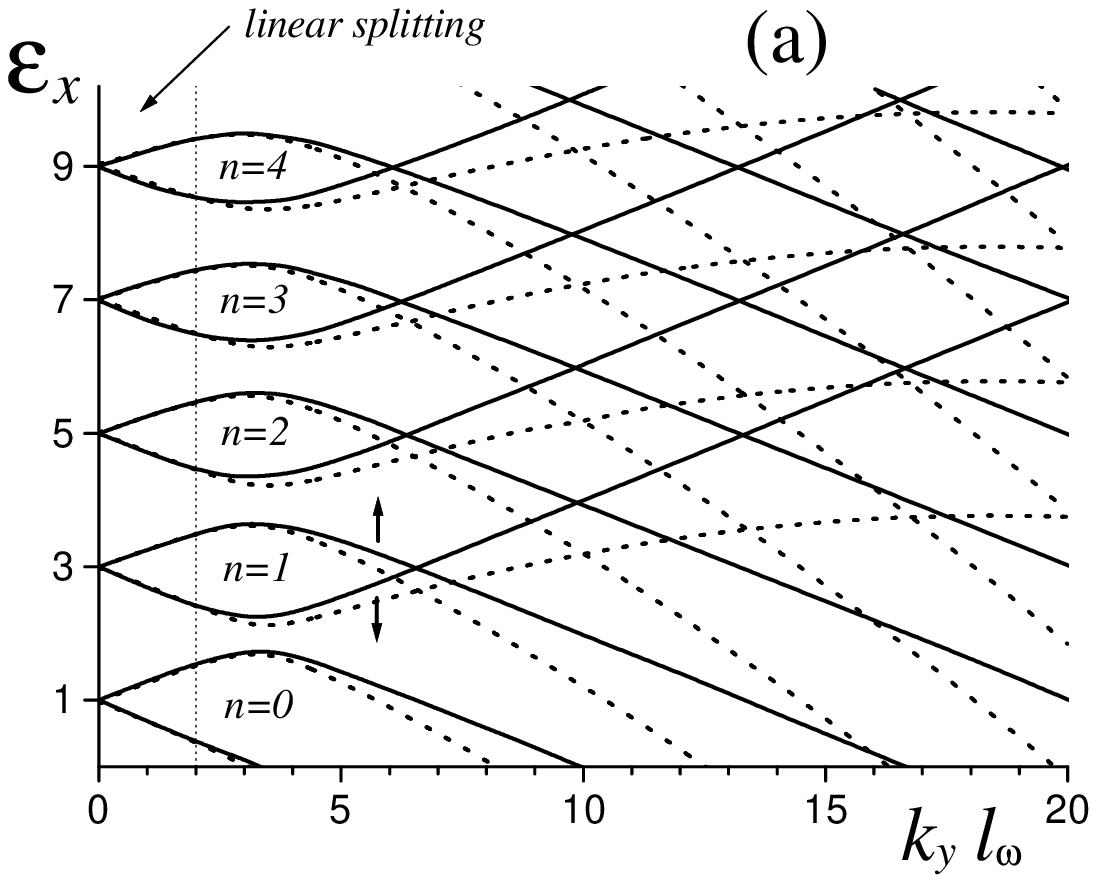,width=3.00in,height=2.00in}
\end{center}
\end{figure}
\vspace{-6ex}
\begin{figure}
\begin{center}
\epsfig{file=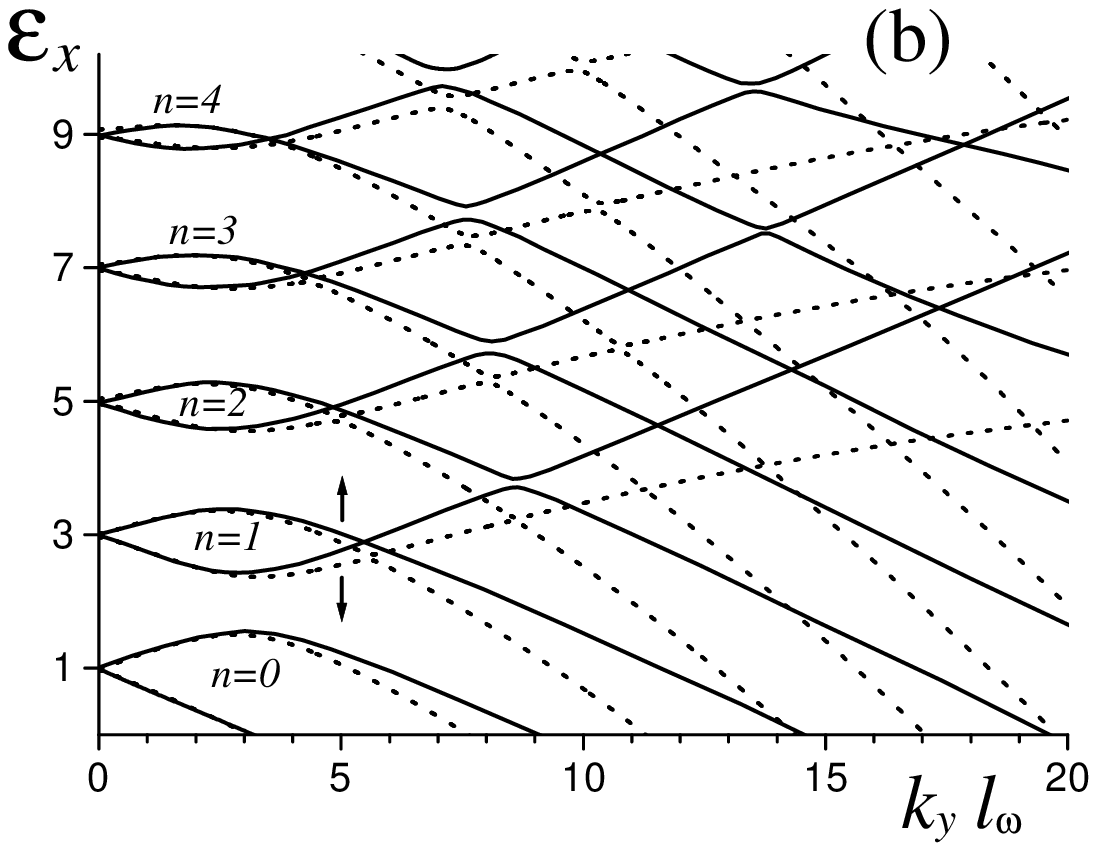,width=3.00in,height=2.00in}
\caption{The transverse energy $\varepsilon_x$ vs. $k_yl_\omega$ for finite
$\alpha$-coupling ($l_\omega/l_\alpha=0.3$): (a) $l_\omega/l_\beta=0$; (b)
$l_\omega/l_\beta=0.1$. Solid and dotted lines correspond to zero
($l_\omega/l_B=0$) and finite ($l_\omega/l_B=0.3$) magnetic field
respectively.}
\label{fig1}
\end{center}
\end{figure}

The application of a weak ($l_\omega/l_B \lesssim 1$) perpendicular
magnetic field bends all the energy curves downwards by an amount
$\propto k_y^2$ [cf.  solid and dotted lines in Fig.~\ref{fig1}(a)]. This
behaviour is consistent with Eq.~(\ref{E0}). We note that a weak magnetic
field has only a small effect on the dispersion law to the left of the
anticrossing region, i.e. for $k_yl_\omega \lesssim 3$. For strong
magnetic fields ($l_\omega/l_B \gtrsim 10$), when the distance between
Landau levels is very large, no anticrossing effects due to the SO
interaction can be seen.

>From Figs.~\ref{fig1}(a,b) it is seen that switching on the
$\beta$-coupling {\it enhances} the anticrossing of ``neighbouring'' energy
branches. Moreover, the strength of the anticrossing now depends on $n$ and
grows with $n$. Interestingly, this effect {\it reduces} the linear energy
splitting $\Delta\varepsilon_x \propto k_y$, in contrast to the expectation
that an additional mechanism of the SO interaction should intensify the
splitting rather than suppress it. What actually happens is that the
$\beta$-coupling, as well as the $\alpha$-coupling, gives a contribution to
the hybridisation of neighbouring electron states~\cite{MB}. As a result,
the hybridisation becomes stronger and leads to the more pronounced
anticrossing and effectively to the {\it suppression} of the energy
splitting. This effect indicates the independent nature of $\beta$-coupling
and its irreducibility to the $\alpha$-coupling. Owing to the enhanced
interstate hybridisation caused by the $\beta$-coupling, the anticrossing
of energy branches in Fig.~\ref{fig1}(b) can be seen in a wider region of
$k_yl_\omega$ up to $k_yl_\omega \approx 13 - 14$. A weak magnetic field
modifies the spectrum in Fig.~\ref{fig1}(b) in basically the same way as it
does in Fig.~\ref{fig1}(a) [cf. solid and dotted lines in
Fig.~\ref{fig1}(b)].

The electron eigenstates that were discussed above can be proven to obey
the fundamental {\it current-conservation identity}~\cite{MB,Baranger} so
that a current can travel adiabatically in any of these states without
scattering into any other. This property allows the low temperature
(ballistic) conductance $G$ of a Q1DES to be calculated directly from the
energy spectrum by relating it to the number $M$ of forward propagating
electron modes at a given Fermi energy $\varepsilon_F$ via simple Landauer
formula~\cite{Landauer}: $G=(e^2/h)M(\varepsilon_F)$.

The most interesting effects on $G$ are obtained for strong
$\alpha$-coupling when $l_\omega/l_\alpha \gtrsim 1.4$. Here the
anticrossing (non-monotonic) portion of curves $\varepsilon_x(k_y)$ in
Fig.~\ref{fig1}(a) comes so close to the $y$-axis that the longitudinal
term $(k_yl_\omega)^2$ in the total subband energy
$\varepsilon=\varepsilon_x+(k_yl_\omega)^2$ does not disguise completely
the original non-monotonicity of $\varepsilon_x(k_y)$. As a result, we see
a small {\it non-monotonic} portion (``bump'') on all the energy curves
$\varepsilon(k_y)$ in Fig.~\ref{fig2}(a) [see a magnified bump for the
lowest level $n=0$ in Fig.~\ref{fig2}(b)]. Remarkably, these bumps give
rise to {\it three} propagating electron modes as opposed to just one
created by any monotonic interval of the spectrum.  Furthermore, two of
these modes [the two leftmost black cicles in Fig~\ref{fig2}(b)] ``mirror''
each other in the sense that they have nearly the same spatial wave
functions but oppositely directed longitudinal group velocities
$v_y=\hbar^{-1}(\partial\varepsilon/\partial k_y)$. It is therefore likely
that weak elastic scattering between the forward and backward propagating
modes may cause directed localisation~\cite{Barnes} and the twin modes will
not contribute to the net current. However, in a sufficiently clean system
the existence of such unusual modes could give rise to sharp ($\sim 0.1$
meV wide) periodic peaks in the dependence $G(\varepsilon_F)$
[Fig.~\ref{fig2}(c)].

A second manifestation of the $\alpha$-coupling in Fig.~\ref{fig2}(c)
is a shift of the conductance quantisation steps to lower energies in
comparison with the case of zero SO interaction (cf. solid and
dotted lines). This effect is caused by energy branches that go downwards
in the region of the linear energy splitting [see Fig.~\ref{fig1}(a)] and
therefore lower the energy of a band edge.

\begin{figure}
\begin{center}
\epsfig{file=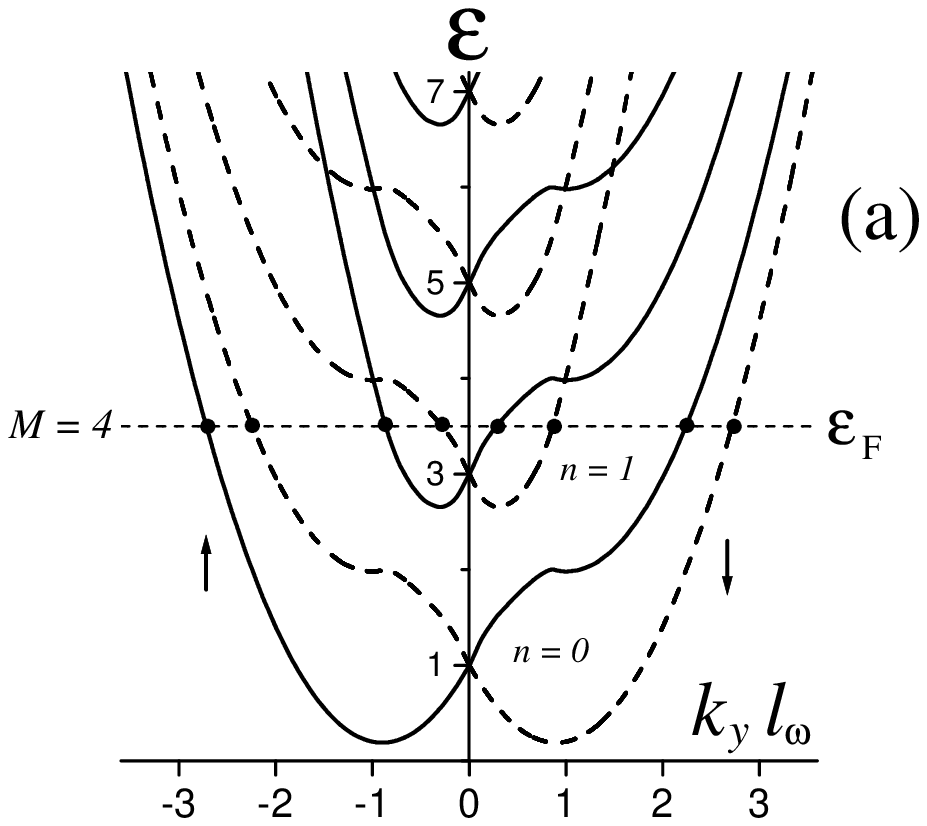,width=3.00in,height=2.00in}
\end{center}
\end{figure}
\vspace{-5ex}
\begin{figure}
\begin{center}
\epsfig{file=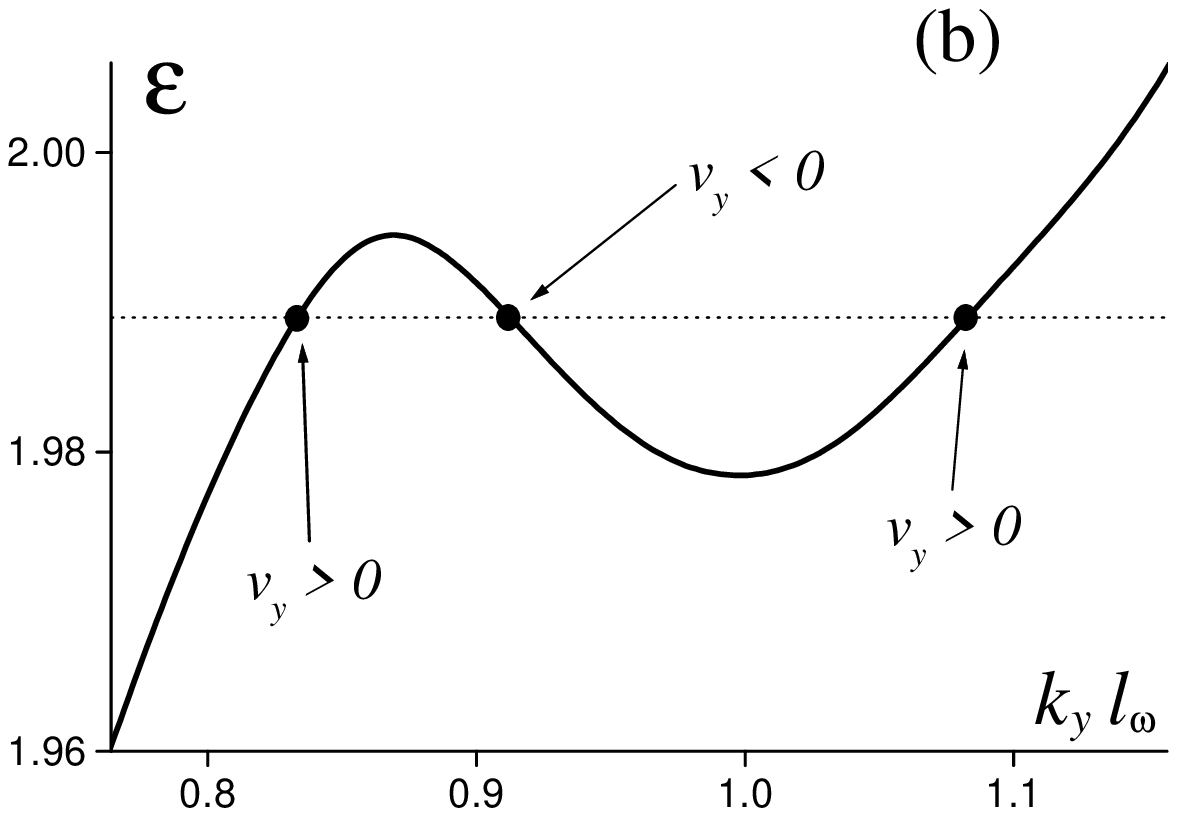,width=2.00in,height=1.50in}
\end{center}
\end{figure}
\vspace{-5ex}
\begin{figure}
\begin{center}
\epsfig{file=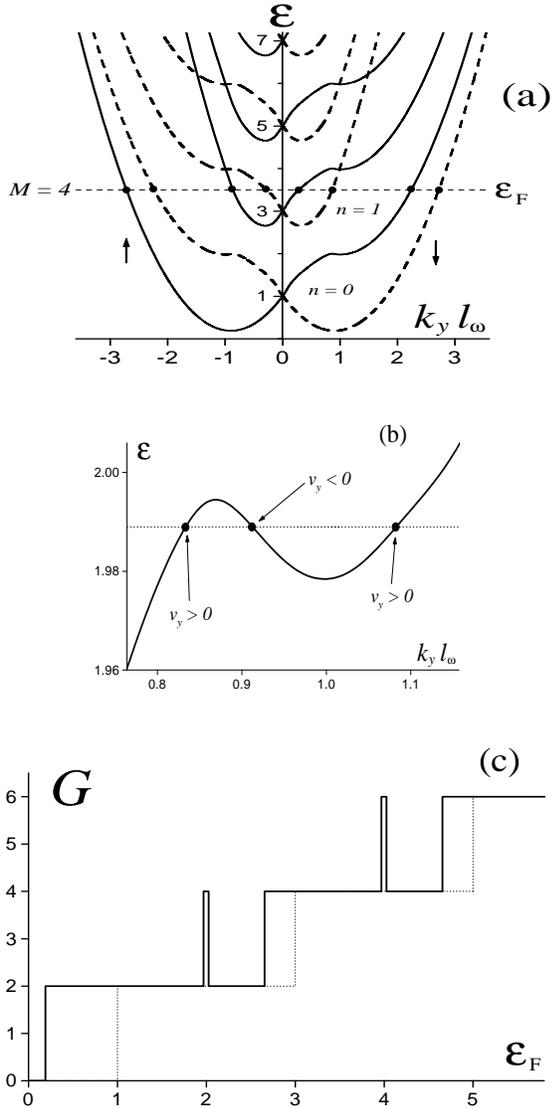,width=3.00in,height=2.00in} \caption{The
dimensionless subband energy $\varepsilon$ vs. $k_yl_\omega$
[figure (a)] and the conductance $G$ vs. the Fermi energy
$\varepsilon_F$ [figure (c)] for $l_\omega/l_\alpha=1.8$,
$l_\omega/l_\beta=l_\omega/l_B=0$. Figure (b) shows a magnified
bump on an $n=0$ energy curve.} \label{fig2}
\end{center}
\end{figure}

In Fig.~\ref{fig1}(b) we saw that switching on the
$\beta$-coupling reduces the energy splitting created by the
$\alpha$-coupling. As applied to the subband energy
$\varepsilon(k_y)$ this means suppression of the non-monotonic
bumps and eventually quenching the peak-like structure in
$G(\varepsilon_F)$. For example, for $l_\omega/l_\beta=0.2$ only
one (the lowest) bump in Fig.~\ref{fig2}(a) survives and hence
only the first peak in $G(\varepsilon_F)$ can potentially be
observed. The existence of the single peak (or just a few peaks)
could be a clear experimental indication of the presence of the
$\beta$-coupling in the system.

In contrast to $\beta$-coupling, a weak perpendicular magnetic field
emphasises the conductance features caused by the $\alpha$-coupling.
Indeed, a {\it negative} effective potential $\propto k_y^2$ due to a
magnetic field [see Eq.~(\ref{E0}) and Figs.~\ref{fig1}(a,b)] compenstates
partially to the contribution of the longitudinal energy $(k_yl_\omega)^2$
to the total subband energy $\varepsilon=\varepsilon_x+(k_yl_\omega)^2$.
Hence the non-monotonic portions of the transverse energy spectrum
$\varepsilon_x$ in Figs.~\ref{fig1}(a,b) are now more important in forming
$\varepsilon$ than they were in zero magnetic field. As a result, the
amplitude (height) of the bumps in Fig.~\ref{fig2}(a) will increase and the
conductance peaks in Fig.~\ref{fig2}(c) become wider (2-3 times). As
regards the peaks destroyed by the $\beta$-coupling, they reappear one by
one starting from the lowest as the magnetic field is being increased.

In conclusion, we have revealed features of the energy spectrum of
electrons and low temperature conductance arising from the specifics of the
spin-orbit interaction in quasi-one-dimensional electron systems:
non-monotonic wave vector dependence of subband energies, anticrossings
between subbands, additional subband minima, and sharp peaks in the
ballistic conductance.

AVM thanks the ORS, COT, and Corpus Christi College for financial
support. CHWB thanks the EPSRC for financial support.


\end{multicols}
\end{document}